\documentstyle[aps,amsmath,amssymb]{revtex}  
\draft  
\begin{document}  
\title{Vortex nucleation and quadrupole deformation of a rotating 
Bose-Einstein condensate}  

\author{M.~Kr\"amer$^a$, L.~Pitaevskii$^{a,b}$, S.~Stringari$^a$
and F.~Zambelli$^a$}  

\address{$^a$Dipartimento di Fisica, Universit\`a di Trento,}  
\address{and Istituto Nazionale per la Fisica della Materia, I-38050
Povo, Italy}  
\address{$^b$Kapitza Institute for Physical Problems, ul. Kosygina 2,
117334 Moscow, Russia}  

\date{\today}  
\maketitle  
\begin{abstract}  
Using a semi-analytic model based on the Thomas-Fermi approximation, we
investigate the relevance of the quadrupole deformation of a trapped
Bose-Einstein condensate for the nucleation of quantized vortices.  
For sufficiently high angular
velocities $\Omega$ of the trap, the tendency of the system to exhibit
spontaneous deformation is shown to lower the barrier which inhibits
the nucleation of vortices at smaller $\Omega$. 
The corresponding value of the critical angular velocity $\Omega_c$ is
calculated as a function of the  deformation of the trap and
of the chemical potential. The theoretical predictions for
$\Omega_c$ refer to the case of a sudden switch-on of the deformed
rotating trap and they are compared with recent experimental data.
\end{abstract}  

 
\begin{section}{Introduction}
\label{intro}
The problem of the nucleation of quantized vortices in Bose
superfluids has been the object of recent experimental work  with
dilute gases in rotating traps  \cite{ens1,mit1,ens2,mit2,ox}.
It now emerges clearly that the mechanism of nucleation depends
crucially on the actual shape of the trap as well as on how the
rotation is switched on.
A first approach consists of a sudden switch-on of the deformation and
rotation of the external potential which generates a
non-equilibrium configuration \cite{ens1,mit1,ens2,mit2}. 
For sufficiently large values of the angular velocity of the rotating 
potential one observes the  nucleation of one or more vortices.
In a second approach either the rotation or the deformation of the
trap are switched on slowly \cite{ens2,ox}. 
In this way, the system 
is in conditions of equilibrium  up to some critical velocity
above which it exhibits a dynamical instability giving rise 
to the nucleation of vortices.
In the case of a sudden switch-on of the deformation and
rotation of the trap, 
the observed critical angular velocity turns out to be close to the
value associated with the instability of the surface oscillations of
the condensate. This occurs  at  the angular 
velocity \cite{muntsa,DS}
\begin{equation}
\omega_{\rm cr}(\ell) = \frac{\omega({\ell})}{\ell}\;,
\label{omegacr}
\end{equation}
where $\omega(\ell)/2\pi$ is the frequency of the surface oscillation
with  angular momentum $\ell$ and we have assumed that the deformation
of the trap is negligibly small.
The actual value of $\ell$ fixing the critical
angular velocity depends explicitly on the shape of the
rotating potential and, to some extent, can be selected in the
experiment \cite{mit2,ENS_april}.
Eq. (\ref{omegacr}) corresponds to the famous Landau criterion for
superfluidity applied to the problem of rotations.
In the Thomas-Fermi limit the critical
angular velocity takes the simple form
\cite{S96} $\omega_{\rm cr}=\omega_{\perp}/\sqrt{\ell}$, where
$\omega_{\perp}$ is the transverse frequency of the harmonic
confinement.
The experimental evidence that Eq. (\ref{omegacr}) provides a good
estimate for the
critical angular velocity for the nucleation of quantized vortices
points out the crucial role played by the shape
deformation of the condensate, in agreement with the theoretical
considerations developed in
\cite{DS,isoshima,castin,feder99,tsubota}. This is further
confirmed by the direct observation \cite{ens2,ox,ENS_april} of strong
deformations occurring temporarily during the process of nucleation.

The purpose of this paper is to gain further insight into the mechanism
underlying vortex nucleation when the deformation and rotation of the
trap are switched on suddenly. 
So far, the explicit relation between the deformation instability and the vortex creation
was not enough clear.
In the absence of shape
deformation the nucleation process is inhibited by the occurrence of a
barrier located near the surface of the condensate. 
Using the Thomas-Fermi (TF) approximation to the Gross-Pitaevskii (GP)
theory at zero temperature \cite{rmp} we will show that this
barrier is lowered by the explicit inclusion of shape deformations and
eventually disappears at sufficiently high angular velocities, making it
possible for the vortex to nucleate. 
We will consider the simple but relevant case of quadrupole
deformations which has already been
the object of systematic experimental studies \cite{ens1,ens2,mit2,ox}.
\end{section}

\begin{section}{Vortices in axisymmetric configurations}
\label{axisymmetric} 

A quantized vortex is characterized by the appearance of a velocity
field associated with a non-vanishing, quantized circulation. 
If we assume that the vortex can be
described by a straight line
the quantization of circulation takes the simple form
\begin{equation}
\hbox{\boldmath $\nabla$}\times{\mathbf v}_{\rm vortex} =
\frac{2\pi\hbar}{m}\,\delta^{(2)}({\mathbf
 r}-{\mathbf d})\hat{z}
\label{curl}
\end{equation}
for a vortex located at distance $d\equiv|{\mathbf d}|$ from the
$z$-axis.
The general solution of (\ref{curl}) can be written in the form
\begin{equation}
{\mathbf v}_{\rm vortex} = \hbox{\boldmath $\nabla$}
\left(\varphi_{\mathbf d} + S\right)
\label{phid}
\end{equation}
where $\varphi_{\mathbf d}$ is the azimuthal angle around
the vortex line at position ${\mathbf d}$ and S is a 
single-valued function which gives
rise to an irrotational component of the velocity field.  The irrotational
component may be important in the case of vortices displaced from the
symmetry axis and its inclusion permits to optimize the energy cost
associated with the vortex line \cite{noteimage}.
Considering a straight vortex line is a first important assumption
that we introduce in our description \cite{note1}.

The inclusion of vorticity is accompanied by the appearance of
angular momentum and by an energy cost. 
Under suitable conditions the system may nevertheless like to acquire
the vortical configuration. This happens if there is a total energy
gain in the rotating frame
where the system is described by the Hamiltonian
\begin{equation}
H(\Omega) = H -\Omega L_z\,.
\label{Homega}
\end{equation}
Here $H$ is the Hamiltonian in the laboratory frame, $L_z$ is the
angular momentum, and $\Omega$ is the angular velocity of the trap
around the $z$-axis.

The simplest way to calculate the angular momentum associated with a
displaced vortex is to assume axi-symmetric trapping and to work
with the TF-approximation. In this limit the size of the vortex core
is small compared to the radius of the condensate
so that one can use the vortex-free expression
$n({\mathbf r})=\mu [1-(r_{\perp}/R_{\perp})^2-(z/Z)^2]/g$ for the density
profile of the condensate, where $r_{\perp}^2=x^2+y^2$ and 
$g=4\pi a \hbar^2/m$ is the coupling constant, fixed by the positive
scattering length $a$.
Then, one can write the angular momentum in the form
\begin{equation}
L_z=m\!\int\!\!dz\!\int\!\!dr_{\perp}r_{\perp} n(r_{\perp},z)
\oint\!\!{\mathbf v}_{\rm vortex}\cdot d{\mathbf l}\,,
\end{equation}
where the line integral is taken along a circle of radius $r_{\perp}$.
Use of Stokes' theorem gives the result \cite{fetter1}
\begin{equation}
L_z(d/R_{\perp})=
N\hbar\left[1-\left(\frac{d}{R_{\perp}}\right)^2\right]^{5/2}\,,
\label{Lzd}
\end{equation}
where $d$ is the distance of the vortex line from the symmetry axis.
Eq. (\ref{Lzd}) shows that the angular momentum per particle
is reduced from the value $\hbar$ as soon as the vortex is displaced
from the center.

To calculate the energy cost associated with the vortex line one
should in  principle start out with Eq. (\ref{phid}) and optimize the
choice for the phase $S$.
In order to obtain a simplified description we will make use of the expression
\begin{equation}
E_v(d/R_{\perp},\mu)=
E_v(d=0,\mu)\left[1-\left(\frac{d}{R_{\perp}}\right)^2\right]^{3/2}\,,
\label{Evd}
\end{equation}
which generalizes the TF-result \cite{lundh}
\begin{equation}
E_v(d=0,\mu) = \frac{4\pi n_0}{3}\frac{\hbar^2}{m}Z
\,\log{\left(\frac{0.671 R_{\perp}}{\xi_0}\right)}
=
N\hbar\omega_{\perp}\frac{5}{4}\frac{\hbar\omega_{\perp}}{\mu}
\log{\left(1.342\frac{\mu}{\hbar\omega_{\perp}}\right)}
\,.
\label{Ev}
\end{equation}
for the energy of an axi-symmetric vortex. 
Here, $Z$ is the TF-radius in $z$-direction and 
$\xi_0$ is the healing length calculated with the central density $n_0$.
The $d$-dependence contained in (\ref{Evd}) is simply understood by
noting that the factor
$4n_0 Z\left[1-(d/R_{\perp})^2\right]^{3/2}\!/3$
corresponds to the column density $\int\!dz\,n(d,z)$ evaluated
with the TF-approximation. Expression (\ref{Evd}) is expected to be correct
within logarithmic accuracy (see also \cite{fetter2} and references
therein). It could be improved by including an explicit
$d$-dependence in the healing length inside the logarithm, in order
to account for the density dependence of the size of the vortex core.

Eqs. (\ref{Lzd}) and (\ref{Evd}) show that when $d=R_{\perp}$ neither
angular momentum nor excitation energy is carried by the system.
Of course these estimates, being
derived with the TF-approximation, are not accurate if we go
too close to the border.
If the surface of the condensate is described by a more realistic density 
profile both the angular momentum and the energy are nevertheless expected
to vanish when the vortex line is sufficiently far outside the bulk region.
Within the simplifying
assumptions made above we can conclude that the
configuration with $d=R_{\perp}$ corresponds to the absence of
vortices, while the transition from $d=R_{\perp}$ to $d=0$ describes the path
nucleating the vortex. The energy associated with each point of this
path in the rotating frame is simply calculated 
using the Hamiltonian (\ref{Homega}). It is given by \cite{fetter1,fetter2} 
\begin{equation}
E_v(d/R_{\perp},\Omega,\mu) = 
E_v(d=0,\mu)\left[1-\left(\frac{d}{R_{\perp}}\right)^2\right]^{3/2}-
\Omega N\hbar\left[1-\left(\frac{d}{R_{\perp}}\right)^2\right]^{5/2}\,.
\label{Evdomega}
\end{equation}
Some interesting features emerge from Eq. (\ref{Evdomega}). First
one finds that the occurrence
of a vortex at $d=0$ is energetically favorable
for angular velocities satisfying the
condition $\Omega \ge \Omega_v(\mu) = E_v(d=0,\mu)/N\hbar$.
This is the well known criterion for the so called thermodynamic
stability of the vortex.
It is worth noticing that in the TF-limit one should
have $\Omega_v(\mu)/\omega_{\perp}<< 1$.
In fact, using the result $\mu= gn_0$ for the chemical potential
and the expression $\mu =m\omega_{\perp}^2R_{\perp}^2/2$, one can write 
\begin{equation}
\frac{\Omega_v}{\omega_{\perp}}=\frac{5}{2}
\left(\frac{a_{\perp}}{R_{\perp}}\right)^2
\log{\left(0.671\frac{R_{\perp}}{\xi_0}\right)}\;,
\label{omegavomega}
\end{equation}
which tends to zero when $R_{\perp}\gg a_{\perp}$.
Here, $a_{\perp}=\sqrt{\hbar/m\omega_{\perp}}$ is the radial oscillator length.
In the actual experiments the ratio (\ref{omegavomega}) is not very small. 
For example, in the case of Ref. \cite{ens1} $\Omega_v\simeq
0.35\,\omega_{\perp}$.

A second interesting result following from (\ref{Evdomega}) concerns
the behaviour of the vortex line for small displacement $d$. 
By expanding $E_v(d/R_{\perp},\Omega,\mu)$ one finds
that the vortex solution at $d=0$ is fully unstable
if $\Omega \le 3\,\Omega_v(\mu)/5$, while it is metastable (local minimum)
if $3\,\Omega_v(\mu)/5\le\Omega\le\Omega_v(\mu)$ \cite{fetter1,fetter2}.

Another important consequence of (\ref{Evdomega}) is the
appearance of a barrier.  Even if $\Omega\ge\Omega_v(\mu)$ and hence
if $E_v(d/R_{\perp},\Omega,\mu)$ is negative at $d=0$, the curve
(\ref{Evdomega}) exhibits a maximum at intermediate values of $d$
 between $0$ and $R_{\perp}$ (see Fig. \ref{fig1}).
The position $d_B$ and height $E_B$ of the barrier are given by the equations
\begin{equation}
\left(\frac{d_B}{R_{\perp}}\right)^2=
1-{3 \over 5}{\Omega_v(\mu)\over \Omega}
\label{equation}
\end{equation}
and
\begin{equation}
E_B =
{2\over 5}E_v(d=0,\mu)
\left[1-\left(\frac{d_B}{R_{\perp}}\right)^2\right]^{3/2} =
{2\over 5}E_v(d=0,\mu)
\left({3 \over 5}{\Omega_v(\mu)\over \Omega}\right)^{3/2}
\label{EB}
\end{equation}
showing that the height of the barrier becomes smaller and smaller as
$\Omega$ increases, but never disappears. 
Since crossing the barrier costs a macroscopic amount of energy, the
system will never be able to overcome it and the vortex cannot be nucleated.

It is finally interesting to rewrite the energy of the vortical
configuration as a function of angular momentum rather than of the
vortex displacement.
This yields the expression
\begin{equation}
E_v(L_z,\Omega,\mu) = 
N\hbar\left[\Omega_v(\mu)\left(\frac{L_z}{N\hbar}\right)^{3/5} -
\Omega \frac{L_z}{N\hbar}\right]\,.
\label{EvLomega}
\end{equation}
Equation (\ref{EvLomega}) emphasizes the fact that the nucleation of
the vortex is associated with an increase of angular momentum 
from zero (no vortex) to $N\hbar$ (one centered vortex),
accompanied by an initial energy increase (barrier) and a
subsequent monotonous energy decrease.
In this form the TF-result can be
usefully compared with alternative approaches based on microscopic
calculations of the vortex energy.
\end{section}

\begin{section}{Role of quadrupole deformations} 

In the previous section we have shown that in order to nucleate the
vortex the system has to overcome a barrier associated with a macroscopic
energy cost. In the following sections we will show that the barrier
disappears at sufficiently high angular velocities if we allow the
system to take a quadrupolar deformation.
The physical mechanism is especially clear for an axisymmetric
trap. Above a given angular velocity,
fixed by (\ref{omegacr}) with $\ell =2$, the system becomes
energetically unstable
against the excitation of quadrupole oscillations. This instability
causes a continuous symmetry breaking
which can favor the occurrence of alternative paths for the
nucleation of the vortex. The appearance
of this intermediate quadrupole deformations is now
strongly supported by experimental evidence \cite{ens2,ox,ENS_april} and
has also been observed in simulations of the time-dependent
Gross-Pitaevskii equation \cite{tsubota}.

In order to describe properly the effects of the
quadrupole deformation we introduce, in addition to the
vortical field (\ref{curl}), an irrotational quadrupolar velocity
field given by
\begin{equation}
{\mathbf v}_{Q} = \alpha \hbox{\boldmath $\nabla$}(xy)\,,
\label{virr}
\end{equation}
where  $\alpha$ is a parameter. Note that ${\mathbf v}_Q$ is the
velocity field in the laboratory frame expressed in terms of the
coordinates of the rotating frame.
The form (\ref{virr}) is suggested by the quadrupolar class of
irrotational solutions exhibited by the time-dependent
Gross-Pitaevskii equation in the rotating frame \cite{alessio}. 
In Ref.\cite{alessio} it has been shown that these solutions
are associated with a quadrupole deformation of the
density described by the deformation parameter
\begin{equation}
\delta = \frac{\langle y^2-x^2\rangle}{\langle y^2+x^2\rangle}\,.
\label{delta}
\end{equation}
The parameters $\delta$ and $\alpha$ characterize the quadrupole
degrees of freedom that we are including in our picture. In the
following, in order to provide a simplified description, we will fix a
relationship between these two parameters by
requiring that the quadrupole velocity field satisfies the condition:
\begin{equation}
\hbox{\boldmath $\nabla$}\cdot
\left[n({\mathbf r})\left({\mathbf v}_Q - \hbox{\boldmath $\Omega$} \times 
{\mathbf r}\right)\right] =0\;,
\label{vQstationary}
\end{equation}
where $\hbox{\boldmath $\Omega$}=\Omega\hat{z}$.
As a consequence of the equation of continuity,
this condition implies that in the rotating frame the density
of the gas is stationary except for the motion of the vortex core which however
involves the change of the density at small length scales.
Eq.(\ref{vQstationary})
yields the relationship $\alpha = -\Omega \delta$ \cite{alessio} which
selects a natural class of paths that will be considered in the
present investigation.
In the presence of the quadrupole velocity field (\ref{virr}), 
the energy of the condensate
in the rotating frame can then be expressed only
in terms of the deformation parameter $\delta$. Using the formalism
of \cite{alessio} one finds the expression:
\begin{equation}
E_Q(\delta,\bar{\Omega},\varepsilon,\mu) =N\mu
\left[\frac{2}{7}\frac{1-\varepsilon\delta-
\bar{\Omega}^2\delta^2}{\sqrt{1-\varepsilon^2}\sqrt{1-\delta^2}}+
\frac{3}{7}\right]\,.
\label{EQ}
\end{equation}
Here, $\varepsilon$ describes the deformation of the
trapping potential in the $x$-$y$ plane
\begin{equation}
V_{\rm ext}({\mathbf r})=\frac{m}{2}\left[(1+\varepsilon)\,\omega_{\perp}^2
x^2+(1-\varepsilon)\,\omega_{\perp}^2y^2
+ \omega_z^2z^2\right]\,,
\label{epsilon_def}
\end{equation}
where $\omega_{\perp}$ is an average transverse oscillator
frequency
\begin{equation}
\omega_{\perp}^2={\omega_x^2+\omega_y^2\over 2}\,,
\end{equation}
and $\bar{\Omega}=\Omega/\omega_{\perp}$ is the angular velocity
of the trap expressed in units of $\omega_{\perp}$.
Note that it is crucial to distinguish between the transverse trap
deformation  $\varepsilon$ and the quadrupole deformation $\delta$ of
the condensate.
In Eq. (\ref{EQ}) we have neglected the change of the central
density caused by the velocity field (\ref{virr}). This assumption will
be used throughout the paper
\cite{notaalessio}.

It is useful to expand Eq. (\ref{EQ}) as a function of $\delta$ in the
case of axisymmetric trapping ($\varepsilon =0$). One finds the result
\begin{equation}  
E_Q(\delta,\bar{\Omega},\varepsilon=0,\mu)\simeq N\mu\left[{5\over 7} +
\delta^2\left(\frac{1}{7}(1-2\bar{\Omega}^2)\right)
+{\cal O}(\delta^3)\right]\,.
\label{expansion2}  
\end{equation}  
which explicitly shows that
for $\Omega > \omega_{\perp}/\sqrt{2}$ the symmetric configuration 
($\delta=0$) is energetically unstable against the
occurrence of quadrupole deformations.
As we will see in the next section this instability is crucial
for the nucleation of the vortex.

Our aim now is to calculate the energy in the rotating frame combining the
effects of the vortex and of the quadrupole deformation.
The parameter $\delta$ thereby emerges as a natural degree of freedom 
that can be used to optimize the path towards vortex nucleation.
In the simultaneous presence of the vortex and of quadrupole
deformation the energy of the system is not however
simply given by the sum of Eqs. (\ref{Evdomega}) and (\ref{EQ}).
In fact, on the one hand the kinetic energy contains the
additional crossed term $m\int\!d{\mathbf r}\,[{\mathbf v}_{\rm
vortex}\cdot {\mathbf v}_{Q}]\,n({\mathbf r})$.
On the other hand we have to take into account
that the vortex energy (\ref{Evd}) and the angular momentum (\ref{Lzd}) were
calculated
for an axisymmetric trap and in the absence of quadrupole deformation.
In the presence of a deformed trap the vortex energy in the laboratory
frame (\ref{Evd}) is simply generalized to
\begin{equation}
E_v(d/R_x,\varepsilon,\mu)=
E_v(d=0,\varepsilon,\mu)\left[1-\left(\frac{d}{R_x}\right)^2\right]^{3/2}\;,
\label{Evdnew}
\end{equation}
where
$R_x$ is the TF-radius of the cloud along $x$ and 
we have explicitly positioned the vortex on the $x$-axis \cite{orientation}. 
It is important to note that the first equality in (\ref{Ev}) also
holds for $\varepsilon\neq 0$. In this case the energy
$E_v(d=0,\varepsilon,\mu)$ can be written in the form
\begin{equation}
E_v(d=0,\varepsilon,\mu)=
N\hbar\omega_{\perp}\frac{5}{4}\frac{\hbar\omega_{\perp}}{\mu}
\sqrt{1-\varepsilon^2}\log{\left(1.342\frac{\mu}{\hbar\omega_{\perp}}\right)}
\;,  
\label{Evnew}
\end{equation}
which points out its dependence on the trap deformation. 
    
Finally, it is possible to also evaluate in a simple
way the sum of the crossed term $m\int\!d{\mathbf r}\,[{\mathbf v}_{\rm
vortex}\cdot {\mathbf v}_{Q}]\,n({\mathbf r})$ entering the kinetic energy
and of the angular momentum term
$-m\,\hbox{\boldmath $\Omega$}\cdot\int\!d{\mathbf r}\,\left[{\mathbf r}
\times{\mathbf v}_{\rm vortex}\right]n({\mathbf r})$ due to the
vortex. One finds
\begin{equation}
E_3(d/R_x,\delta,\bar{\Omega})=
m\!\int\!\!d{\mathbf r}\;{\mathbf v}_{\rm vortex}\cdot\left[{\mathbf v}_{Q}
-\hbox{\boldmath $\Omega$}\times{\mathbf r}\right]n({\mathbf r})
=-N\hbar\omega_{\perp}\bar{\Omega}
\,\sqrt{1-\delta^2}\,\left[1-\left(\frac{d}{R_x}\right)^2\right]^{5/2}
\label{crossed}
\end{equation}
which is consistent with (\ref{Lzd}) in the case of a
symmetric condensate ($\delta=0$). In deriving result (\ref{crossed})
we have used the condition (\ref{vQstationary})
for the quadrupole velocity field and we have integrated by parts using the
expression (\ref{phid}) for ${\mathbf v}_{\rm vortex}$.
Note that the irrotational component $\nabla S$ of ${\mathbf v}_{\rm vortex}$ 
does not contribute to (\ref{crossed}).
\end{section}

\begin{section}{Critical angular velocity for vortex nucleation}  

The total energy of a quadrupolar deformed condensate in the presence
of a vortex is given by the sum of Eqs. (\ref{Evdnew}), (\ref{EQ}) and 
(\ref{crossed}):
\begin{equation}
E_{\rm tot}(d/R_x,\delta,\bar{\Omega},\varepsilon,\mu)=
E_v(d/R_x,\varepsilon,\mu)+E_Q(\delta,\bar{\Omega},\varepsilon,\mu)+
E_3(d/R_x,\delta,\bar{\Omega})\,,
\label{totalenergy}
\end{equation}
where we have explicitly indicated the dependence of the three
contributions on the various physical parameters.
The values of the angular velocity $\Omega$, the trap anisotropy
$\varepsilon$, the average transverse oscillator frequency 
$\omega_{\perp}$, and the chemical potential $\mu$ are fixed by
the experimental conditions. Hence the degrees of freedom of the system
with which one can play in order to identify the optimal path for 
vortex nucleation are the vortex displacement $d$ and the condensate
deformation $\delta$.
Note that each of the three energy contributions has a different dependence
on the chemical potential resulting in a non-trivial dependence of the
critical angular velocity on the relevant parameters of the system
(see Eq. (\ref{Om_c_analytic}) below). Result (\ref{totalenergy})
generalizes the one given in Ref. \cite{fetter1}, which holds only in the
limit of small $\bar{\Omega}$ where $\delta\sim\varepsilon$. 

We assume that the system is initially in the state $d/R_x=1$, $\delta=0$.
This assumption adequately describes an experiment in which $\Omega$
and $\varepsilon$ are switched on suddenly.
In this case, the condensate is initially axisymmetric ($\delta=0$)
and vortex-free ($d/R_x=1$).
Of course this configuration is not stationary and will evolve in
time. In the following
we will make use of energetic considerations in order to explore the
possible paths followed by the system towards the nucleation of the vortex
line. These paths should be
associated with a monotonous decrease of the energy.
In Figs. \ref{fig2} and \ref{fig3}  we have plotted the
energy surface $E_{\rm tot}$ for two different values of the
angular velocity $\bar{\Omega}$ and fixed $\varepsilon$ and
$\mu/\hbar\omega_{\perp}$.
In both cases the angular velocity was chosen high enough to make the
vortex state a global energy minimum.
This minimum is surrounded by an energy ridge which, for $\delta=0$,
forms a barrier between the initial state ($d/R_x=1$) and the vortex
state ($d=0$), as discussed in section \ref{axisymmetric}.
Moreover, the energy ridge exhibits a saddle point at non-zero
deformation $\delta$.
The height of this saddle depends on $\bar{\Omega}$,
$\varepsilon$, and $\mu/\hbar\omega_{\perp}$.
In Fig. \ref{fig2} the energy at the saddle point is higher than the
energy of the initial state and the ridge can not be surpassed.
However, at higher angular velocities $\bar{\Omega}$ the situation changes.
In Fig. \ref{fig3} the saddle lies lower than the initial state.
Hence in this case the system can bypass the barrier by crossing the saddle.
The corresponding path is always associated with the occurrence of a
strong intermediate deformation of the condensate.

The critical angular velocity for the nucleation of vortices naturally
emerges as the angular velocity $\bar{\Omega}_c$ at which the energy
on the saddle point $((d/R_x)_{\rm sp},\delta_{\rm sp})$ is the same as the
energy of the initial state $(d/R_x=1,\delta=0)$:
\begin{equation}  
E((d/R_x)_{\rm sp},\delta_{\rm sp},\bar{\Omega}_c,\varepsilon,\mu)=
E(d/R_x=1,\delta=0,\bar{\Omega}_c,\varepsilon,\mu)\,.
\label{def_om_c}
\end{equation}  
It is worth mentioning that crossing the saddle point is not the
only possibility for the system to lower its energy.
In fact Figs. \ref{fig2} and \ref{fig3} show the existence of
stationary deformed
vortex-free states which can be reached starting from the initial
state. These are the states predicted in \cite{alessio} and
experimentally studied in \cite{ens2,ox} through an adiabatic increase of
either $\Omega$ or $\varepsilon$ instead of doing a rapid switch-on.
The energy ridge separates these configurations from the vortex state.
Still, under certain conditions this stationary vortex-free state becomes
dynamically unstable and a vortex can be nucleated starting out from
it \cite{ens2,ox,castin}.
The study of this type of vortex nucleation is beyond the scope of the
present  paper.

The actual value of the critical angular velocity $\bar{\Omega}_c$ for
vortex nucleation depends on the parameters $\varepsilon$ and
$\mu/\hbar\omega_{\perp}$.
Fig. \ref{fig4} shows the dependence of $\bar{\Omega}_c$ on $\varepsilon$ for
different choices of $\mu/\hbar\omega_{\perp}$.
To lowest order in $\varepsilon$ and $\hbar\omega_{\perp}/\mu$
the dependence is given by
\begin{equation}  
\bar{\Omega}_{\rm c}(\varepsilon)\!-\!\frac{1}{\sqrt{2}}
\:\approx\:\frac{1}{\sqrt{2}}\left[\frac{(A\eta)^{1/2}}{4}
-\frac{\varepsilon}{(A\eta)^{1/4}}\right]\,,
\label{Om_c_analytic}
\end{equation}
where 
$$\eta=\left[\log{\left(1.342\,\frac{\mu}
{\hbar\omega_{\perp}}\right)}\right]^{5/2}
\left(\frac{\hbar\omega_{\perp}}{\mu}\right)^{7/2}$$ 
and $A=2^{-5/4}21\sqrt{3}$.
This formula shows that the relevant parameters of the expansion are
$\eta^{1/2}$ and $\varepsilon/\eta^{1/4}$. Fig. \ref{fig4} demonstrates
that it is applicable also at rather small values of the chemical
potential.

At $\varepsilon=0$ the vortex nucleation, according to the present
scenario, is possible only at angular
velocities slightly higher than the value $\omega_{\perp}/\sqrt{2}$.
For non-vanishing $\varepsilon$ the preferable configuration will be
always deformed even for small values of $\bar{\Omega}$ where
$\delta$ depends linearly on $\varepsilon$.
At higher $\bar{\Omega}$ the condensate gains energy by increasing its
deformation in a nonlinear way (see \cite{alessio}).
Eq. (\ref{Om_c_analytic})  and Fig. \ref{fig4} shows that for non-vanishing 
$\varepsilon$
the saddle point on the energy ridge can be surpassed at
angular velocities smaller than $\omega_{\perp}/\sqrt{2}$.

The above scenario seems to be in reasonable agreement with
experiments. In particular, evidence for critical angular velocities
smaller than $\omega_{\perp}/\sqrt{2}$ occurring when $\varepsilon\neq
0$ has emerged from the experiments reported in \cite{ENS_april,mit2}.
A more systematic comparison with the expected dependence of
$\bar{\Omega}_c$ on the trap deformation $\varepsilon$ and on the
chemical potential $\mu$ would be crucial in order to assess the
validity of the model \cite{oxfordnote}. 
\end{section}

\begin{section}{Stability of a vortex-configuration against
quadrupole deformation}  
\label{stability}

Once the energy barrier is bypassed, the vortex moves to the center of
the trap ($d/R_x=0$) where the energy has a minimum. In an axisymmetric trap
($\varepsilon=0$) this configuration will be in general stable against
the formation of quadrupole deformations of the condensate unless the
angular velocity $\Omega$ of the trap becomes too large. The
criterion for instability is easily obtained by studying the
$\delta$-dependence of the energy of the system in the presence of a single
quantized vortex located at $d/R_x=0$.
Considering the total energy (\ref{totalenergy}) we find 
\begin{equation}  
E_{\rm tot}(d/R_x=0,\delta,\bar{\Omega},\varepsilon=0,\mu)\simeq
E_{\rm tot}(d/R_x=0,\delta=0,\bar{\Omega},\varepsilon=0,\mu)+
\delta^2N\mu\left(\frac{1}{7}(1-2\bar{\Omega}^2)+
{\bar{\Omega}\over 2}\frac{\hbar\omega_{\perp}}{\mu}\right)+
{\cal O}(\delta^3)\;.
\label{expansion1}  
\end{equation}   
Comparing Eqs. (\ref{expansion1}) with the analog expression (\ref{expansion2})
holding in the absence of the vortex line one observes that in the
presence of the vortex the instability against quadrupole deformation
occurs at a higher angular velocity given by
\begin{equation}  
\Omega=\omega_{\perp}\left(\frac{1}{\sqrt{2}} +
\frac{7}{8}\frac{\hbar\omega_{\perp}}{\mu}\right)\;. 
\label{cr2}  
\end{equation}
If the angular velocity is smaller than (\ref{cr2}) the vortex
is stable in the axisymmetric configuration while
at higher angular velocities the system prefers to deform, giving rise
to new stationary configurations.
The critical angular velocity (\ref{cr2}) can also be obtained
by applying the Landau criterion (\ref{omegacr}) to the quadrupole 
collective frequencies in the presence of a quantized vortex.
These frequencies were calculated in \cite{francesca} using a sum rule 
approach.
For the $m=\pm 2$ quadrupole frequencies the result reads
\begin{equation}  
\omega_{\pm 2} =\omega_{\perp}\sqrt{2} \pm \frac{\Delta}{2}\;,
\end{equation}  
where 
\begin{equation}
\Delta =
\omega_{\perp}\left(\frac{7}{2}\frac{\hbar\omega_{\perp}}{\mu}\right)\,
\end{equation} 
is the frequency splitting between the two modes.
Applying the condition (\ref{omegacr})
to the $m=+2$ mode one can immediately reproduce result (\ref{cr2}) for the
onset of the quadrupole instability in the presence of the
quantized vortex.

In an anisotropic trap ($\varepsilon\ne 0$), the stable vortex state
will generally be associated with a non-zero deformation $\delta$ of
the condensate. It is interesting to note that $\delta$ increases with
the angular velocity of the trap and easily exceeds $\varepsilon$ (see
Figs. \ref{fig2} and \ref{fig3}). This behaviour is analogous to the
properties of the deformed stationary states in the absence of
vortices \cite{alessio}. 
\end{section}

\begin{section}{Final comments and conclusions}

In this paper we have developed a semi-analytic model to describe the
nucleation of a quantized vortex induced by the sudden switch-on of the
trap deformation and rotation. The trap was assumed to be of
quadrupolar shape. 
The model includes the distance of the vortex line
from the principal axis and the quadrupole deformation of the condensate as
the key degrees of freedom of the system. We have shown that, for
sufficiently high angular velocities, the instability exhibited by a
condensate with respect to surface quadrupole deformations gives rise to a
path for vortex nucleation. The energy diagram is characterized by the
occurrence of a saddle point whose height can favor or inhibit the
nucleation, depending on the value of the angular velocity, the
deformation of the trap and the chemical potential. The scenario
for the nucleation emerging from the present approach agrees with the
experimental results recently obtained in \cite{ens1,ens2,mit2,ox,ENS_april}
as well as with the predictions based on the numerical solution of the
time-dependent Gross-Pitaevskii equation \cite{tsubota}.
In particular, the nucleation is always associated with an intermediate
configuration exhibiting a significant quadrupole deformation of the
condensate, even in the presence of a very small deformation of the
trap. Furthermore the nucleation is favored by increasing the values
of the trap deformation and the value of the chemical potential.

The energy diagram employed to calculate the critical angular velocity
was based on the Thomas-Fermi approximation to the vortex
energy. This approximation is expected to become worse and worse as
the vortex line approaches the surface region \cite{franco96,lundh}. 
Since the main mechanism of nucleation takes place near the surface, we
expect that the quantitative predictions 
concerning the dependence of the critical angular velocity on the trap
deformation and on the chemical potential would be improved
by using a microscopic evaluation of the vortex energy, beyond the
Thomas-Fermi approximation. This should include, in particular,
the density dependence of the size of the vortex core, as well as
the proper inclusion of image vortex effects. Work in this direction
is in progress.

Useful discussions with Y.~Castin, F.~Dalfovo, J.~Dalibard, D.~Feder,
E.~Hodby, O.~Marag\`o and A.~Recati are acknowledged.
\end{section}
\begin{figure}
\input{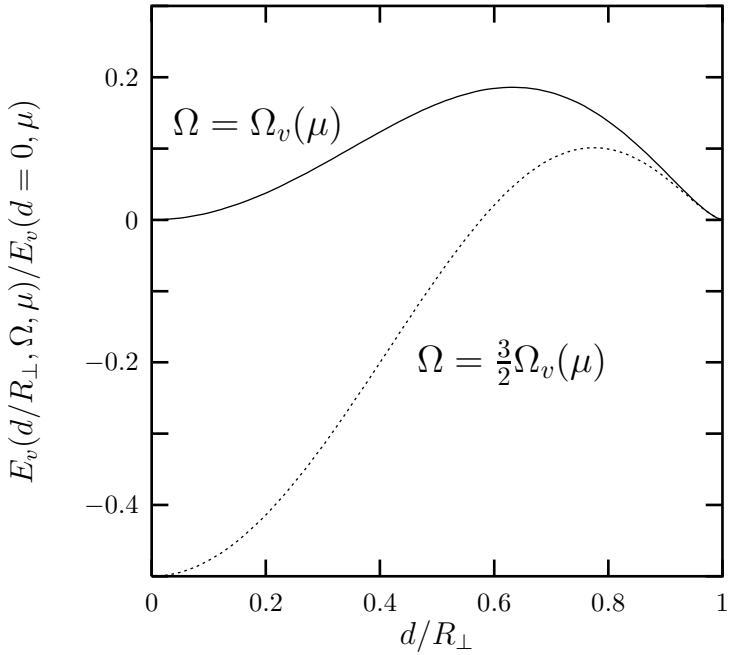}
\caption{
Vortex excitation energy in the rotating frame (\ref{Evdomega}) for 
an axisymmetric
configuration as a function of the reduced vortex displacement
$d/R_{\perp}$ from the center. The curves refer to two different choices 
for the angular velocity of the trap: $\Omega=\Omega_v(\mu)$ (solid line), and
$\Omega=3\,\Omega_v(\mu)/2$ (dotted line), where
$\Omega_v(\mu)=E_v(d=0,\mu)/N\hbar$. The initial vortex-free state 
corresponds to $d/R_{\perp}=1$. For $\Omega>\Omega_v(\mu)$, 
the state with a vortex at the center ($d/R_{\perp}=0$) is preferable.
However, in this configuration the nucleation of the vortex is inhibited by
a barrier separating the vortex-free state ($d/R_{\perp}=1$) from the energetically favored
vortex state ($d/R_{\perp}=0$).
}
\label{fig1}
\end{figure}
\newpage
\begin{figure}
\input{fig2_map064.tex}
\caption{Below the critical angular velocity of vortex nucleation:
the plot shows the dependence of the total energy
(\ref{totalenergy}) minus the energy of the initial non-deformed
vortex-free state
$E(d/R_x=1,\delta=0)$ on the quadrupolar shape deformation $\delta$ and
on the vortex displacement $d/R_x$ from the center. Energy is given in
units of $N\hbar\omega_{\perp}$. 
The dashed line corresponds to $E_{\rm tot}-E(d/R_x=1,\delta=0)=0$,
while the solid curve refers to  $E_{\rm
tot}-E(d/R_x=1,\delta=0)=0.015N\hbar\omega_{\perp}$.
This plot has been obtained by setting $\varepsilon =0.04$, 
$\mu=10\,\hbar\omega_{\perp}$ and $\Omega =0.64\omega_{\perp}$.
The initial state is indicated
with $\oplus$, while $\otimes$ corresponds to the energetically
preferable centered vortex state.
The barrier $\oslash$ inhibits vortex nucleation in a non-deforming
condensate ($\delta=0$).
The saddle point $\odot$ lies lower than the barrier $\oslash$.
However, at the chosen $\Omega$ the energy on the saddle is still
higher than the one of the initial state $\oplus$. 
Note that the preferable vortex state is associated with a shape
deformation $\delta>\varepsilon$ (see section
\ref{stability}).
Note also the existence of a favorable deformed and vortex-free state
labeled by $\bigtriangledown$ \protect\cite{alessio}.}
\label{fig2}
\end{figure}
\newpage
\begin{figure}
\input{fig3_map07.tex}
\caption{
Above the critical angular velocity of vortex nucleation:
the plot shows the dependence of the total energy
(\ref{totalenergy}) minus the energy of the initial non-deformed
vortex-free state $E(d/R_x=1,\delta=0)$ on the quadrupolar shape
deformation $\delta$ and on the vortex displacement $d/R_x$ from the
center. Energy is given in units of $N\hbar\omega_{\perp}$. 
The dashed line corresponds to $E_{\rm tot}-E(d/R_x=1,\delta=0)=0$,
while the solid curve refers to  $E_{\rm
tot}-E(d/R_x=1,\delta=0)=0.015N\hbar\omega_{\perp}$.
In this plot $\varepsilon =0.04$, $\mu=10\,\hbar\omega_{\perp}$, as in
Fig. \ref{fig2}, and $\Omega =0.7\omega_{\perp}$. Important states are
indicated as in Fig. \ref{fig2}. At the chosen $\Omega$, the saddle
$\odot$ lies lower than the initial state $\oplus$ allowing the system
to bypass the barrier $\oslash$ by taking a quadrupolar deformation
$\delta$ and reach the preferable vortex state $\otimes$. 
Note also the existence of a favorable deformed and vortex-free state
labeled by $\bigtriangledown$ \protect\cite{alessio}.}
\label{fig3}
\end{figure}
\newpage
\begin{figure}
\input{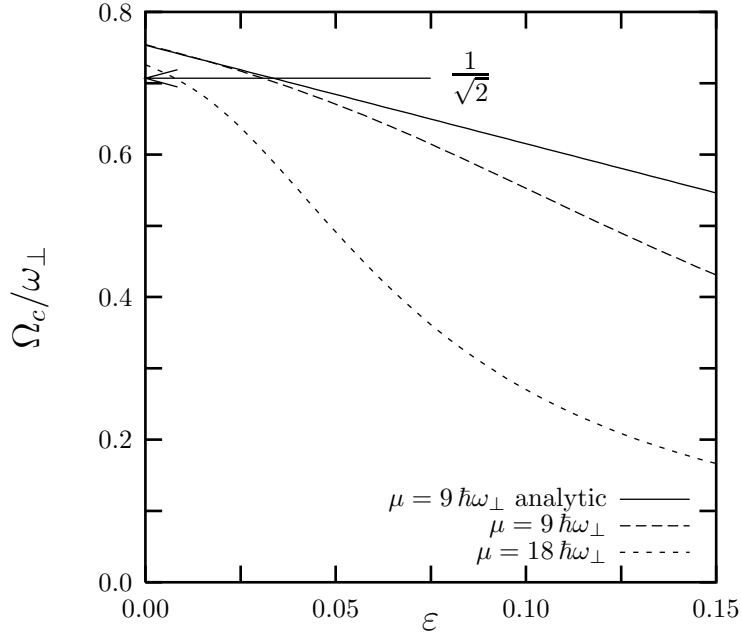}
\caption{
Critical angular velocity of vortex nucleation in units of
$\omega_{\perp}$ as a function of the trap deformation $\varepsilon$. 
The long dashed and short dashed curves correspond to the numerical 
calculation satisfying condition (\ref{def_om_c}) for
$\mu=9\,\hbar\omega_{\perp}$ and $\mu=18\,\hbar\omega_{\perp}$
respectively, while the solid line is the analytic prediction
(\ref{Om_c_analytic}) evaluated with $\mu=9\,\hbar\omega_{\perp}$. The
arrow indicates the angular velocity at which the quadrupole surface mode
becomes unstable in the case $\varepsilon=0$.
The value $\mu=9\,\hbar\omega_{\perp}$ is close to the experimental
setting  of \protect\cite{ens1}.}
\label{fig4}
\end{figure}


\begin{references}
\bibitem{notaalessio}
Imposing that the energy (\ref{EQ}) be stationary with respect to $\delta$ yields the solutions
derived in \cite{alessio} apart from small corrections due to the changes of the central density
not accounted for in the present formalism.
\bibitem{orientation}
Our results do not change if $d/R_x$ is replaced by $d/R_y$.
We expect that predicting the preferable direction for a vortex to enter the condensate demands the calculation of the vortex energy beyond logarithmic 
accuracy.
\bibitem{oxfordnote} 
A first experimental curve of $\bar{\Omega}_c(\varepsilon)$ in 
qualitative agreement with our predictions has recently been reported
in \cite{ox}. However the measured values of $\bar{\Omega}_c$ are
systematically shifted upwards, probably because the sudden switch-on
condition was not satisfied in this experiment.
\end{references}
\end{document}